# Searching for Kardashev Type III civilisations from High $q$-Value Sources in the LoTSS-DR1 Value-added Catalogue

H. Chen[1,2,3] and M. A. Garrett[2,4]
[1]National Astronomical Observatories, Chinese Academy of Sciences, A20 Datun Road, Chaoyang District, Beijing, 100101, China
[2]Jodrell Bank Centre for Astrophysics (JBCA), Department of Physics & Astronomy, Alan Turing Building, The University of Manchester, M13 9PL, UK
[3]University of Chinese Academy of Sciences, 19A Yuquanlu, Beijing, 100049, China
[4]Leiden Observatory, Leiden University, PO Box 9513, 2300 RA Leiden, The Netherlands

**ABSTRACT**
Kardashev Type III civilisations have by definition energy requirements that are likely to generate strong excess emission in the mid-infrared (MIR) that is associated with the waste heat they generate. For those civilisations that capture starlight via Dyson sphere like constructions, the Optical to MIR flux ratio of the host galaxies is also expected to be unusually low. Since a wide range of galaxy types adhere to the infrared-radio correlation (IRC), galaxies hosting Type III civilisations should also strongly deviate from this relation. Radio data can therefore play a crucial role in breaking the degeneracy between the effects of dust obscuration in a galaxy and the signature of an energy-intensive civilisation. We have used the newly released LoTSS-DR1 value-added catalogue to create a sample of 16,367 $z < 0.2$ sources with radio and MIR flux densities, optical photometry and (photometric) redshifts. We calculated the IRC parameter $q_{150\ \mathrm{MHz}}^{22\ \mu\mathrm{m}} = \log(S_{150\ \mathrm{MHz}}/S_{22\ \mu\mathrm{m}})$ and identified 21 sources with exceptionally high values of $q_{150\ \mathrm{MHz}}^{22\ \mu\mathrm{m}}$, an indication of MIR emission enhanced by a factor of $\sim 10$. Out of the 21 high $q$-value sources, 4 sources have very red MIR colors, and appear to have relatively low optical/MIR ratios. Two of the 4 sources are not well known in the literature, they are considered as potential hosts of Type III civilisations. These sources deserve further study and investigation. Finally, we note that extending our analysis to the full LoTSS survey area can place very strong constraints on the incidence of Type III civilisations in the Universe.

## 1 INTRODUCTION

The well-established far-infrared-radio correlation applies to a wide variety of galaxies in the local Universe (see Solarz et al. 2019 and references therein). Similar trends are also seen in the mid-infrared (MIR)-radio domain (Donley et al. 2005; Park et al. 2008). As demonstrated by Garrett (2002) and Elbaz et al. (2002), the correlation continues to apply at cosmological distances and over 7 orders of magnitude in luminosity. The tightness of the Infrared Radio Correlation (hereafter IRC) is usually interpreted as the non-thermal radio emission and the thermal IR emission both being related to processes associated with massive star formation (see Magnelli et al. 2015 for a review). However, the IRC also applies to non-starforming galaxies, including many radio-quiet AGN. Radio-loud AGN (RL AGN; $L_{1.4\ \mathrm{GHz}} \gtrsim 10^{24}$ W Hz$^{-1}$ in the local Universe) deviate from the IRC, presenting very low values of $q$, the latter being defined as the logarithmic ratio between the IR and radio flux densities. The value of $q$ has often been used to disentangle AGN activity from star-forming processes in extragalactic objects, e.g. Condon et al. (2002); Sargent et al. (2012); Delhaize et al. (2017). On the other hand, radio-quiet AGN (RQ AGN), or very young star-forming galaxies (SFGs) with comparably low radio emission may cause elevated values of $q$ (Costagliola et al. 2016; Roussel et al. 2003). So far, a satisfactory physical model that is consistent with all observed constraints and that fully explains the IRC and its outliers is still to be fully developed (Read et al. 2018 and Wang et al. 2019). A better understanding of the astrophysical nature of sources with very high values of $q$ is useful in progressing studies in this field.

In addition to natural astrophysical objects like RQ AGNs and young SFGs possessing high values of $q$, it has been proposed that artificial sources, such as Kardashev Type III civilisations (Kardashev 1964) can also present the same unusual signature Garrett (2015). A Kardashev Type III civilisation is defined as being able to draw on a large fraction of the stellar energy available from its host galaxy ($\sim 4 \times 10^{37}$ W; Kardashev 1964). As suggested by Wright et al. (2014a) and Wright et al. (2014b), Type III civilisations may be identified via their waste heat signature which is expected to peak at MIR wavelengths, assuming peak temperatures of $\sim 100-600$ K (Carrigan 2009). A significant fraction of the optical emission in the galaxy is also expected to be absorbed by artificial structures such as Dyson Spheres or Swarms (Dyson 1960). This leads to the identification of three characteristics a galaxy hosting a Type III civilisation might be expected to have: (1) extremely red MIR colours, (2) high values of $q$, and (3) unusually low optical/IR luminosity ratios.

To characterize the first effect in the MIR, Wright et al. (2014a) developed a formalism to define the fractional starlight re-radiated as alien waste heat ($\gamma$) by fitting the measured Wide-field Infrared Survey Explorer (*WISE*) MIR data to a spectral energy distribution (SED) model for normal spiral galaxies. This formalism reveals the mechanism of how the waste heat from a Type III civilisation will potentially generate excessive MIR emission with very red colours.

Since the radio emission of a galaxy is largely expected to remain unaffected by the presence of a Type III civilisation, such systems should present large values of $q$, well in excess of the average value (Garrett 2015). As suggested by Wright et al. (2014a), with an approximation to a blackbody with temperature T* and the Rayleigh-Jeans limit ($\lambda \gg hc/kT^{*}$), the expected SED of a home galaxy of Type III civilisation can be written as:

$$F_{\lambda} \approx \gamma \left( \frac{T^{*}}{T_{\rm waste}} \right)^{3} F_{\lambda}\,(\gamma = 0)\,. \quad (1)$$

where $T_{\rm waste}$ is the characteristic temperature of the waste heat, and $F_{\lambda}(\gamma = 0)$ is the flux density of a normal galaxy.

For example, taking $T^{*} \sim 6000$ K, and $T_{\rm waste} \sim 100 - 600$ K, and assuming values of $\gamma \sim 0.01 - 0.1$, the MIR flux will be enhanced by a factor of 10−100. If the typical value of $q$ is ~ 0.5, then values in excess of ~ 1.5 to 2.5 might be anticipated.

These estimates are likely to hold for quiescent galaxies with little to no star-formation (Ilbert et al. 2013). Quiescent galaxies account for about half of the stellar mass in the Universe since z ~ 1, so they may be important as hosts of advanced civilisations. However, the Rayleigh-Jeans limit would fail in typical star-forming galaxies, in which the MIR radiation is mostly contributed by cold dust. We quantify the MIR enhancement for such sources in Section 3.1.

Griffith et al. (2015) constructed a sample of 93 sources (the Ĝ sample) with $\gamma > 0.25$ out of ~ 100,000 *WISE* sources with good photometry. Many of these sources has little or no presence in the scientific literature and were proposed as potential advanced Type III civilisations. However Garrett (2015) cross-matched the Ĝ sample with the NRAO/VLA Sky Survey (NVSS) catalog, and found that the vast majority of the sources have $q$ values consistent with the IRC. This suggests that the MIR excess and prominent red colours of the sources was likely to be associated with natural astrophysical processes. It turns out that samples solely based on MIR data produce many false-positives −− star forming galaxies naturally present a MIR excess as a consequence of the reprocessing of optical/UV emission by dust. Indeed, searches for anomalies in astronomical data that target deviations from galaxy scaling laws in general (e.g. outliers in the Tully-Fisher relation, Annis 1999) have similar shortcomings. Radio data can therefore play a crucial role in breaking this degeneracy between the effects of dust obscuration in a galaxy and the signature of an energy-intensive civilisation.

The newly released LOw Frequency ARray (LOFAR; van Haarlem et al. 2013) Two-metre Sky Survey first data release (LoTSS-DR1; Shimwell et al. 2017) value-added catalogue (Shimwell et al. 2019; Williams et al. 2019; Duncan et al. 2019) contains ~ 300,000 radio sources cross-matched with the *WISE* catalogue. This is an excellent resource for the investigation of the nature of sources with high values of $q$ and the identification of potential Kardashev Type III civilisations. Optical measurements are also available for these sources, in particular Panoramic Survey Telescope and Rapid Response System (Pan-STARRS) $i$-band photometry data. In this work, we perform an IRC and colour distribution analysis based on the LoTSS-DR1 value-added catalogue to build a sample of sources which display high values of $q$, extreme red MIR colours, and unusually low optical/MIR flux ratios. It is notable however, our radio-selected sample might be biased to starbursts and AGN with strong radio emission. To calculate $q$-values, the measured LoTSS radio flux is required, therefore host galaxies of Type III civilisations with radio flux below the detection threshold (0.6 mJy) of LoTSS-DR1 would be missed in this methodology. On the other hand, very hot Dyson spheres like the "Matrioshka brain" (Bradbury et al. 2011) could reach a waste heat of ~ 1,000 K, however, since Type III civilisations with $T_{\rm waste} \gtrsim 1{,}000$ K would peak at a wavelength below 3.4 $\mu$m and therefore be excluded in the colour distribution analysis, we only look for type III civilisations with $T_{\rm waste} \sim 100 - 600$ K.

Our approach is complimentary to the earlier work of Wright et al. (2014a,b) and Griffith et al. (2015) but takes advantage of the new large scale radio surveys now being conducted by telescopes such as LOFAR and the other SKA pathfinders/precursors. We introduce the LoTSS-DR1 value-added catalogue in Section 2, followed by a description of our high-$q$ source selection in Section 3. In Section 4, we present MIR colour and optical magnitude analysis to find potential host galaxies of Type III civilisations. A summary and suggestion for future steps are given in Section 5.

## 2 THE LOFAR DR1 VALUE-ADDED CATALOGUE

LoTSS is an ambitious low-frequency radio survey with a wide range of science goals that aims to observe the entire northern sky at frequencies of 120 − 168 MHz with a resolution of ~ 6″. Using the High Band Antenna (HBA) system of LOFAR, the survey reaches a sensitivity of less than 0.1 mJy beam$^{-1}$. The first LoTSS data release covers 424 square degrees over a field centred on the Hobby-Eberly Telescope Dark Energy Experiment (HETDEX; Hill et al. 2008). In a series of papers, Shimwell et al. (2019); Williams et al. (2019); Duncan et al. (2019) built up a LoTSS-DR1 "value-added" catalogue[1] including 318,520 discrete radio sources identified by the Python Blob Detector and Source Finder (PyBDSF; Mohan & Rafferty 2015). The value-added catalogue was cross-matched with the Pan-STARRS 3$\pi$ Steradian survey (Chambers et al. 2016), and the All-sky *WISE* (*All-WISE*) catalogue (Cutri et al. 2021). The source associations and optical and/or IR identifications were done by using a combination of statistical techniques and visual inspections (Williams et al. 2019). Photometric redshift information is given for 94.4% of the LoTSS-DR1 sources (Duncan et al. 2019).

The *All-WISE* catalogue includes photometric data from four MIR channels at 3.4, 4.6, 12, and 22 $\mu$m (W1, W2, W3, and W4) for more than 747 million sources over the full sky. The *WISE* images have a typical point spread function of 6−6.5″ for W1, W2, and 12″ for W3 and W4.

The Pan-STARRS 3$\pi$ Steradian Survey was observed in 5 bands (*grizy*) covering the entire northern sky of ẟ > 30°. The mean 5$\sigma$ point source limiting sensitivities for the 5 bands are 23.3, 23.2, 23.1, 22.3 and 21.4 mag respectively. There is a systematic uncertainty of 7-12 millimag in the photometric calibration depending on the bandpass.

There are 231,716 (73%) sources in the final LoTSS-DR1 value-added catalogue having a WISE and/or Pan-STARRS counterpart, amongst them, 55,694 were detected in WISE W4 and have profile fitted W4 flux density data, which can be used to measure the IRC parameter $q^{22\,\mu m}_{150\,\rm MHz}$ (see Section 3). While the LoTSS value-added catalogue is dominated by extragalactic sources, we cannot rule out the possibility that our sample may be contaminated by a few Galactic objects. However, considering that the HETDEX field is ~ 60° off of the Galactic plane, we do not expect this to be a major problem for our analysis.

[1] The value-added catalogue is available online at https://lofar-surveys.org/releases.html

In this paper, the prefix "ILT" before the sourcenames refers to the LoTSS catalogue.

## 3 SAMPLE OF HIGH $Q$-VALUE SOURCES

### 3.1 Candidate Selection

In Section 1, we use the SED model of a Type III civilisation with Rayleigh-Jeans limit in Wright et al. (2014b) to estimate the enhancement of their MIR emission. It should be noted that these estimates are likely to hold for quiescent galaxies but may not be applicable to all galaxy types. Many quiescent galaxies are early-type galaxies (ETG). As shown by Nyland et al. (2017), ~ 19−50% of ETGs have higher IR-radio ratios than is typical for SFGs. Possible explanations for this include the presence of nascent SF in these systems or specific properties of the galaxy e.g. a bottom-heavy stellar IMF, enhanced cosmic ray escape, weak magnetic fields, or other environmental effects. Looking for sources with $q$-value +1 than normal would filter out most of the high-$q$ ETG outliers, but there may still be false-positives. For star-forming galaxies, however, the majority of the 22 $\mu$m emission is not from stars but from cold dust (~ 10−20 K) and small grains, the Rayleigh-Jeans limit would not be applicable here. According to Figure 15 of Draine & Li (2007), the value of $\nu\,L_\nu$ ($\nu$ = 24 $\mu$m ≈ 22 $\mu$m) in such galaxies is about 3−6% of the total flux. As shown in Figure 1, the value of $\nu\,L_\nu$ ($\nu$ = 22 $\mu$ m)/$L_{total}$ peaks at ~ 0.7 at 160 K. If an advanced civilisation captured all starlight with Dyson spheres before it can be absorbed by dust, $\nu\,L_\nu$ ($\nu$ = 22 $\mu$ m)/$L_{total}$ raises to ~ 10−70% for Dyson spheres with a blackbody temperature of 100−600 K. In other words, these galaxies would see their MIR luminosity enhanced by a factor of 1.7−23.3, resulting in $q_{22\,\mu m}$-values increased by up to ~ +1.4 dex in excess of normal values. For star-forming systems with an initial $\nu\,L_\nu$ ($\nu$ = 22 $\mu$m)/$L_{total}$ = 0.03 and a Dyson sphere with $T_{waste}$ ≲ 100 K or ≳ 400 K, the enhanced $\nu\,L_\nu$ ($\nu$ = 22 $\mu$ m)/$L_{total}$ value would be ≲ 0.3, which is less than 10 times to the original value, these sources may be largely contaminated by AGN.

In this paper, we only look for Type III civilisations with their MIR radiation at least 10 times enlarged, this would allow us to find Type III civilisations in quiescent galaxies with $\gamma$ ~ 0.01 and $T_{waste}$ ~ 100 − 600 K or star-forming galaxies with $T_{waste}$ ~ 100 − 400 K. Galaxies in which the MIR emission is being enhanced by a factor of 10, should present $q$ values raised linearly by +1.

We use the *WISE* 22 $\mu$m and the 120−160 MHz LOFAR measurements to calculate the IRC parameter: $q^{22\,\mu m}_{150\,MHz}$, which is defined as $\log(S_{22\,\mu m}/S_{150\,MHz})$. We restricted ourselves to galaxies with redshifts < 0.2 in order to have good auxiliary data for sources that warranted more detailed investigation. Another consequence of this limited redshift range is that a $k$-correction need not be applied to the data. To check this assumption, we divided the sources into two sub-samples: $z$ < 0.1 and 0.1 < $z$ < 0.2. Out of the ~ 300,000 LoTSS-DR1 sources, there are 5790 and 10577 sources with *WISE* W4 photometric flux data, and redshifts of $z$ < 0.1 and 0.1 < $z$ < 0.2, respectively. The $z$ < 0.1 sources have an average $q^{22\,\mu m}_{150\,MHz}$ value and a standard deviation of ~ 0.49 ± 0.37, while the sources with 0.1 < $z$ < 0.2 have a mean $q^{22\,\mu m}_{150\,MHz}$ of ~ 0.48 ± 0.30. Figure 2 shows the distribution of $q^{22\,\mu m}_{150\,MHz}$ value versus calculated *WISE* 22 $\mu$m luminosity. As shown in Figure 2, $q^{22\,\mu m}_{150\,MHz}$ tends to increase along the 22 $\mu$m luminosity, this trend probably arises from the higher number of starburst galaxies in the more luminous sample, for which a more intense radiation field and hot dust are expected. In addition, starburst galaxies tend to have slightly flatter radio spectra and more high frequency emission at > 5 GHz due to free-free from hot gas. As illustrated by left panel of Figure 2, in the low 22 $\mu$m luminosity and low redshift regime, the fit seems much stronger than any actual trend. To filter out false-positive in low-$z$ and low-$L_{22\,\mu m}$ regime, we plot the median $q_{22}$ value of sources in Yun et al. (2001) calculated by Nyland et al. (2017), and a ± $\log_{10}5$ constraints for normal galaxies, corresponding to 5 times the IR-excess, which applies to most of galaxies in Yun et al. (2001). The median $q_{22}$ value of Yun et al. (2001) sample of 0.99 was defined as the logarithmic ratio between $L_{22\,\mu m}$ and $L_{1.4\,GHz}$ (Nyland et al. 2017), we convert their median $q_{22}$-value to $q^{22\,\mu m}_{150\,MHz}$ assuming a spectral index between $L_{1.4\,GHz}$ and $L_{150\,MHz}$ of -0.755, which is the averaged value for AGN and SFGs measured in Calistro Rivera et al. (2017). As a result, in the scenario of Yun et al. (2001), most galaxies would have a $q^{22\,\mu m}_{150\,MHz}$ value within ~ -0.45 to 0.95, and a median value of ~ 0.25. Figure 3 is a histogram of distance of $q^{22\,\mu m}_{150\,MHz}$ value to the linear fit over $L_{22\,\mu m}$. For clarity, the source count scale is logarithmic. As demonstrated by Figures 2 and 3, the majority (~ 99.9%) of the sources fell below the linear fits +1 and 0.25 + $\log_{10}5$ (= 0.95).

We classify sources with $q^{22\,\mu m}_{150\,MHz}$ larger than the linear fits +1 and higher than 0.25 + $\log_{10}5$ as high $q$-value sources. There are 20 sources selected in the $z$ < 0.1 sample and 1 source in the 0.1 < $z$ < 0.2 sample. We therefore identify these 21 (20 + 1) sources as parent sample of possible Kardashev Type III candidates.

We investigate the physical properties of these high-$q^{22\,\mu m}_{150\,MHz}$ value sources in the following sections.

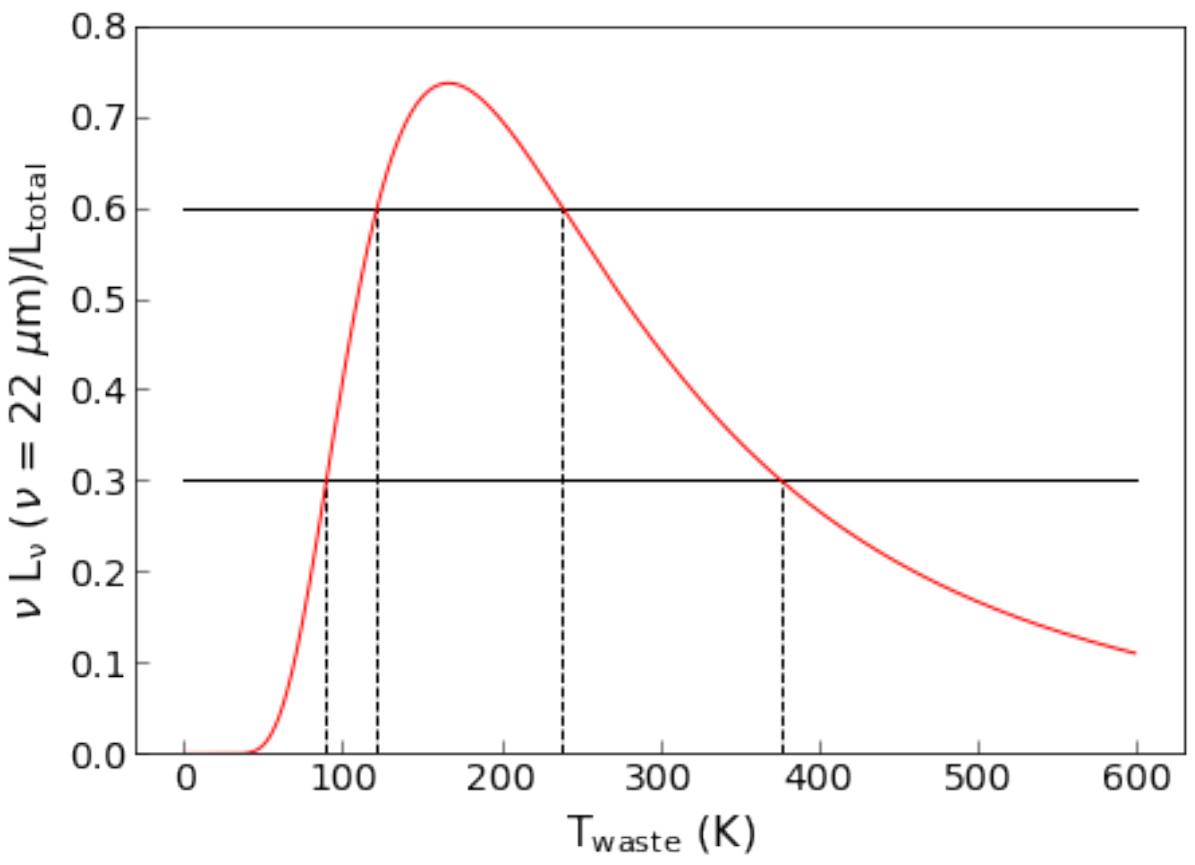

(a)

**Figure 1.** Fractional 22 $\mu$m radiation to the total flux of a blackbody (Dyson sphere), the horizontal lines denote where $L_\nu$ ($\nu$ = 22 $\mu$m)/$L_{total}$ = 0.3 and 0.6, the lowest values required for star-forming galaxies with $\nu\,L_\nu$ ($\nu$ = 22 $\mu$m)/$L_{total}$ = 0.03 and 0.06 to be enlarged by at least 10 times.

### 3.2 Source Classification

We used `SIMBAD`[2] to search within 10″ around the positions of the 21 high-$q^{22\,\mu m}_{150\,MHz}$ value sources looking for object identifiers in the literature. An overview of the source classification is shown in Table 1. The majority (19/21) of the sources are previously identified in the literature. Of the 19 sources with identifiers, 9 of them are identified as AGN, 4 are classified as emission-line galaxies, as well as one composite galaxy showing both star-forming and AGN features, the other 6 sources are confirmed to be galaxies but with no specific type designated. We have calculated the radio luminosity of the sources −− they are all radio-faint with $L_{150\text{ MHz}} \lesssim 10^{23}$ W Hz$^{-1}$. The source properties, including radio and MIR flux densities, radio luminosity, $q^{22\,\mu\text{m}}_{150\text{ MHz}}$ values, and redshift information, as well as the names of the identifiers in the literature for the high-$q^{22\,\mu\text{m}}_{150\text{ MHz}}$ value sources are listed in Table 2.

[2] http://simbad.u-strasbg.fr/simbad/

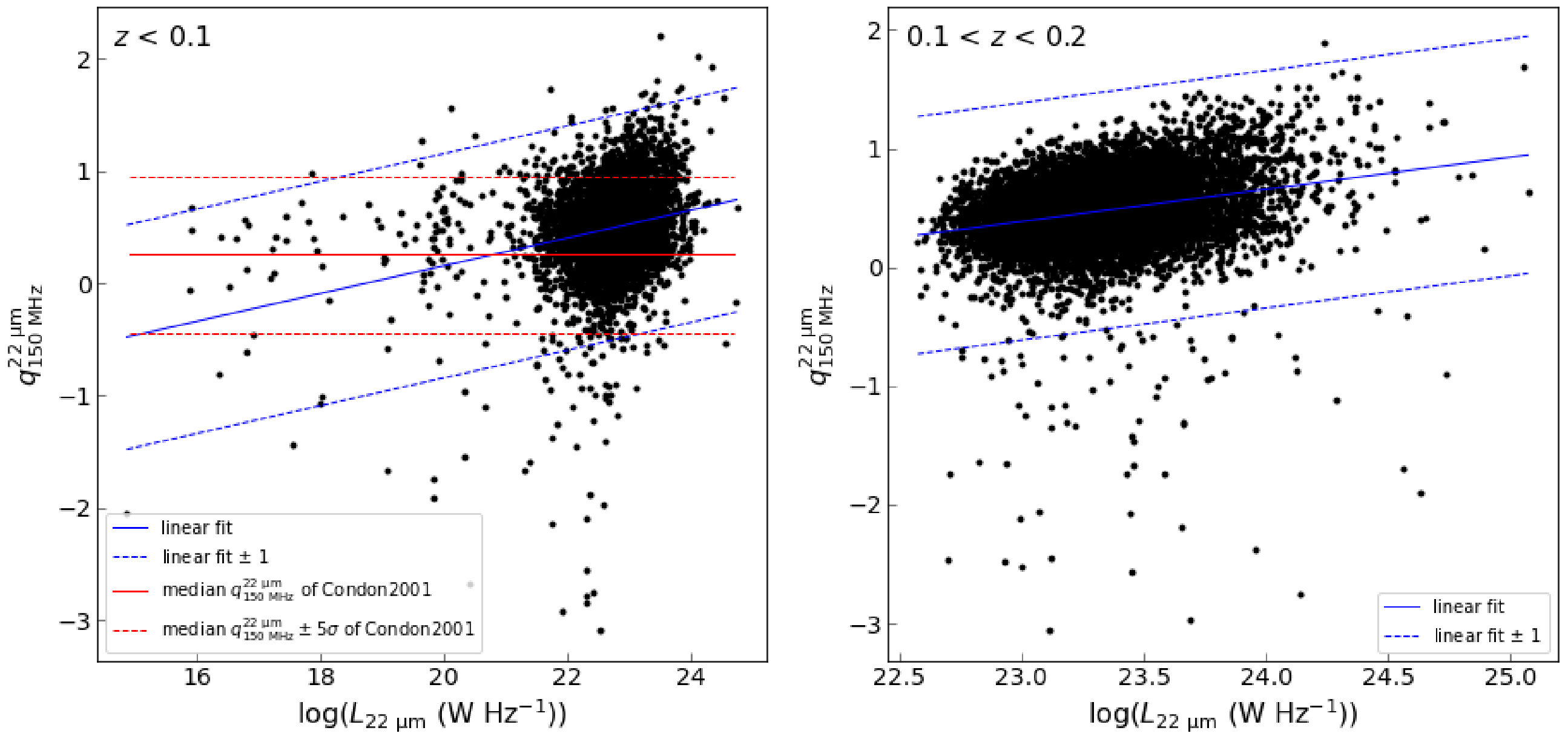

(a) (b)

**Figure 2.** $q^{22\ \mu m}_{150\ \mathrm{MHz}}$ v.s. *WISE* W4 luminosity for sources with (a) $z < 0.1$ and (b) $0.1 < z < 0.2$. The solid and dashed lines denote the least square linear fitting over the sources, the dashed line restrain the region between ± 1 to the fitting line, the solid and dashed red lines define the median $q^{22\ \mu m}_{150\ \mathrm{MHz}}$ value of ∼ 0.25 for sources in Yun et al. (2001), and the 5-time IR-excess/deficit.

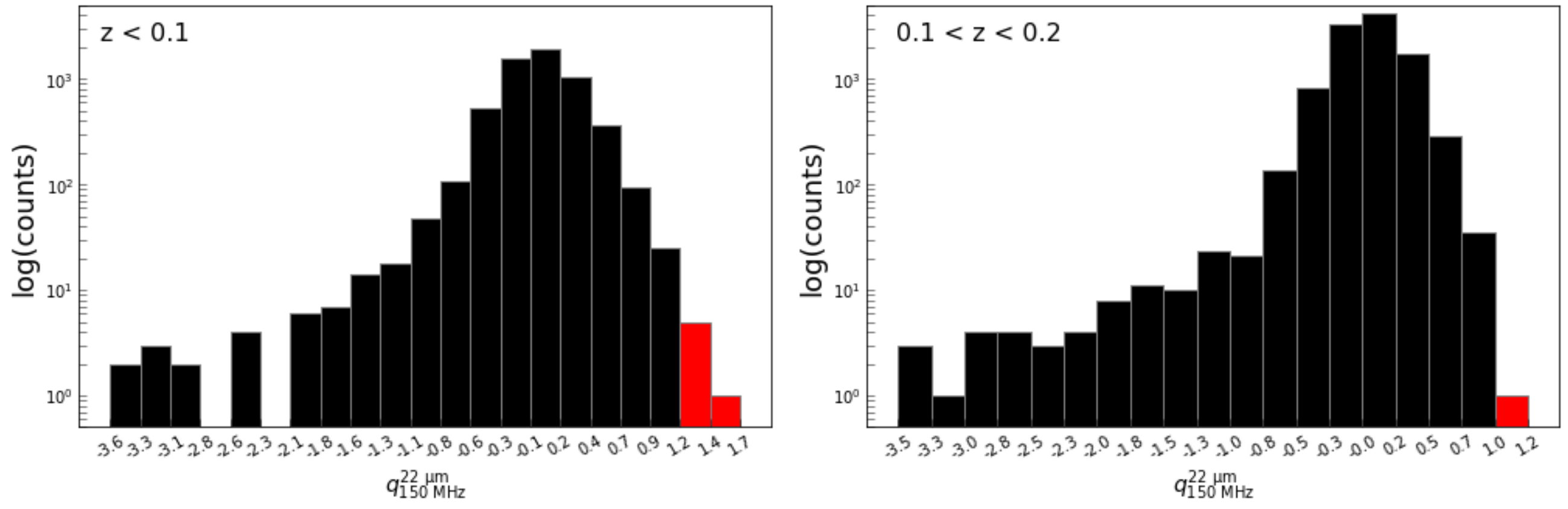

(a) (b)

**Figure 3.** The histogram of distance of $q^{22\ \mu m}_{150\ \mathrm{MHz}}$ value to the linear fit over $L_{22\ \mu m}$ in a logarithmic scale, the red bins represent sources with $q^{22\ \mu m}_{150\ \mathrm{MHz}}$ values larger than the linear fitting +1 and higher than $0.25 + \log_{10}5$.

**Table 1** Numbers of each type of sources with high value of $q^{22\ \mu m}_{150\ \mathrm{MHz}}$.

| Type[1] (total counts) | AGN[2] (9) | | | | | EmG[3] (4) | | Comp (1) | Other (5) | | No identifier (2) | Total Sources |
|---|---|---|---|---|---|---|---|---|---|---|---|---|
| Classification[4] | Sy1 | Sy2 | LIN | AGN | QSO | H2G | EmG | Comp | G | GiP | | |
| Counts | 4 | 2 | 2 | 1 | 1 | 3 | 2 | 1 | 4 | 1 | 2 | 21 |

[1]Counts include all selected sources with redshift up to 0.2.
[2]This sub-sample contains a multi-associated sources: ILT J112637.84+513422.9 (`SIMBAD` identifier: MCG+09-19-078) that it classified both as a Seyfert 1 . Galaxy in (Dong et al. 2012; source ID: J112637.74513423.0) and Seyfert 2 Galaxy in (Toba et al. 2014; source ID: MCG +09-19-078)
[3]This sub-sample contains a multi-associated source, ILT J140956.92+545649.5 (`SIMBAD` identifier: LEDA 2816082), that is classified as an emission-line galaxy in (Shim & Chary 2013; source ID: SDSS 587735696987389965) and an HII galaxy in (Shirazi & Brinchmann 2012; source ID: 1325-52762-412.)
[4]Sy1: Seyfert 1 Galaxy; Sy2: Seyfert 2 Galaxy; LIN: LINER-type AGN; QSO: quasar object; EmG: Emission-line galaxy; H2G: HII galaxy; Comp: Composite galaxy with both star-forming and AGN activity; G: Galaxy; GiP: Galaxy in pair of galaxies.

AGN are the most common type of sources in our high-$q^{22\,\mu\text{m}}_{150\text{ MHz}}$ value sample. The vast majority of AGN are RQ AGN with radio powers that are at least 3 orders of magnitude smaller than the much brighter RL AGN. The radio emission in these systems is potentially generated by several different processes, in addition to non-thermal emission from accretion and the ejection of low-powered jets, accretion disk coronal activity, free-free emission from photo-ionised gas, and star-formation on sub-kpc and kpc scales are also expected to be major sources of radio emission in this class of object (Panessa et al. 2019; Padovani et al. 2015). The majority of RQ AGN will typically follow the IRC but some galaxies such as those in our sub-samples can have high values of $q$ when their IR emission is enhanced by the presence of dust, usually located in the central regions surrounding the AGN. In addition to AGN, there are also 4 high-$q^{22\,\mu\text{m}}_{150\text{ MHz}}$ value sources that are identified as emission-line galaxies which are possibly dominated by star-forming processes. Additionally, there is one source that is classified as a composite galaxies with both AGN and star-forming activity. As suggested by Galvin et al. (2016); Costagliola et al. (2016) and Roussel et al. (2003), radio continuum emission is a delayed tracer of starformation due to the long timescales involved ($\sim 10^7$ years) for electrons to diffuse within galaxies on kpc scales. For galaxies with a very recent onset of starformation, this can lead to higher values of $q$ being measured. There are 5 sources identified as galaxies of one type or another in our sub-samples but two sources (ILT J134649.72+542621.7 and ILT J145757.90+56532) have very little presence in the literature and have not been well studied to date.

### 3.3 Additional $q$-value Distributions

In addition to $q^{22\,\mu\text{m}}_{150\text{ MHz}}$ that is derived from the *WISE* W4 22 $\mu$m and 150 MHz LOFAR flux densities, we can also calculate additional $q$-values for the high-$q$ sub-samples using the W1, W2 and W3 *WISE* MIR bands. Within the 5790 $z < 0.1$ and 10577 $0.1 < z < 0.2$ sources that have *WISE* W4 photometric flux, there are 5775 and 10559 of them that also have *WISE* channel 1, 2 and 3 data available. As suggested by Wright et al. (2014a) and Wright et al. (2014b), the waste heat of a Type III civilisation is expected to peak at 100−600 K, corresponding to an excess MIR emission that peaks between 5−30 $\mu$m. Since the radio flux densities used to measure the values of $q$ are constant for each source, the $q$-value distribution curves for the sources reflect the shape of the source SEDs in the MIR. Figure 4 shows the $q$-values for the high-$q^{22\,\mu\text{m}}_{150\text{ MHz}}$ value sources versus wavelengths of the four *WISE* channels. As shown in Figure 4, our sources generally have an SED that rises towards longer MIR wavelengths, which is similar to the model of a typical spiral galaxy built by Silva et al. (1998) - no explicit features of Type III civilisations can be seen in either sub-sample.

## 4 CANDIDATE TYPE II/III CIVILISATIONS

In this section, we look at the MIR colours and optical properties of sources exhibiting high $q^{22\,\mu\text{m}}_{150\text{ MHz}}$ value (see Section 3.1) to look for potential hosts of Type III civilisations.

### 4.1 MIR Colour Distribution

In order to find sources with extreme MIR colours among our high-$q^{22\,\mu\text{m}}_{150\text{ MHz}}$ sample, we plot the *WISE* w1 - w2 colour distribution over the two sub-samples in Figure 5. The criteria of W1 - W2 > mean + 2$\sigma$ is applied on the sources, however, we note that these sources could be the upper tail of the colour distribution, rather than a new population. As shown in Figure 5a and b, there are 9 high-$q^{22\,\mu\text{m}}_{150\text{ MHz}}$ sources that appear to be outliers in the top-right of each panel −− 8 with $z < 0.1$ and 1 with $0.1 < z < 0.2$. We investigate the optical properties of these sources in the next section.

### 4.2 Optical vs. MIR source characteristics

In Figure 6, we plot the Pan-STARRS $i$-band monochromatic AB magnitude (see Tonry et al. 2012 for details) versus *WISE* W4 luminosity for the two sub-samples, including the 9 extremely red high-$q^{22\,\mu\text{m}}_{150\text{ MHz}}$ sources. Within the 5790 $z < 0.1$ and 10577 $0.1 < z < 0.2$ sources that have *WISE* W4 photometric fluxes, there are 5708 and 10541 of them that also have PanSTARRs $i$-band forced aperture photometry magnitude data available. The Pan-STARRS $i$-band is the most sensitive band of the survey, with a bandpass of 690−819 nm, and a capture cross section of $\sim 8.5$ m$^2 e^-$ photon$^{-1}$ at 1.2 airmass (Tonry et al. 2012).

As shown in Figure 6, the optical magnitude of the sources correlates with increasing 22 $\mu$m flux density, the solid orange lines show the linear fits to 22 $\mu$m over the samples. We select sources with their $M_i$ 5$\sigma$ under the linear fit as the optical-faint sample. We note that these sources could be the optically faint tail of the sources rather than a new population. Of the 9 sources that have relatively high values of $q$ and are extremely red in the MIR, 4 of these are underluminous, all four sources have a redshift of < 0.1.

We present the radio, MIR and optical images of these 4 sources in Figure 7, the field of view (FoV) of each image is 0.5′′. The LoTSS 150 MHz images retrieved from the LoTSS-DR1 image cutout service[3] have a pixel size of 1.5′′. The *WISE* multi-colour images were retrieved from the NASA/IPAC Infrared Science Archive[4] and have a pixel size of 1.25′′. The latter are combined images, summing by the 3.4 (blue), 4.6 (green), and 12 (red) $\mu$m images together. The Pan-STARRS optical images have a pixel size of 0.25′′, and are combined images using the Pan-STARRS y (red), i (green), g (blue) filters −− these images are retrieved from the Pan-STARRS Data Release 1 Image Cutout Server[5]. All the sub-figures in Figure 7 are centered on the target coordinates drawn from the LoTSS DR1 catalogue. As shown in this figure, the sources appear to be point-like and very red in the MIR. These four sources are therefore candidate Type III civilisations.

Among the 4 candidate sources, two sources were classified as natural sources in the literature. ILT J120658.79+524126.5 (`SIMBAD` identifier: SDSS J120658.79+524126.4), was identified as a Seyfert 1 galaxy in (Toba et al. 2014; source ID: SDSS J120658.79+524126.4)

[3] https://lofar-surveys.org/releases.html
[4] https://irsa.ipac.caltech.edu/applications/wise/
[5] http://ps1images.stsci.edu/cgi-bin/ps1cutouts

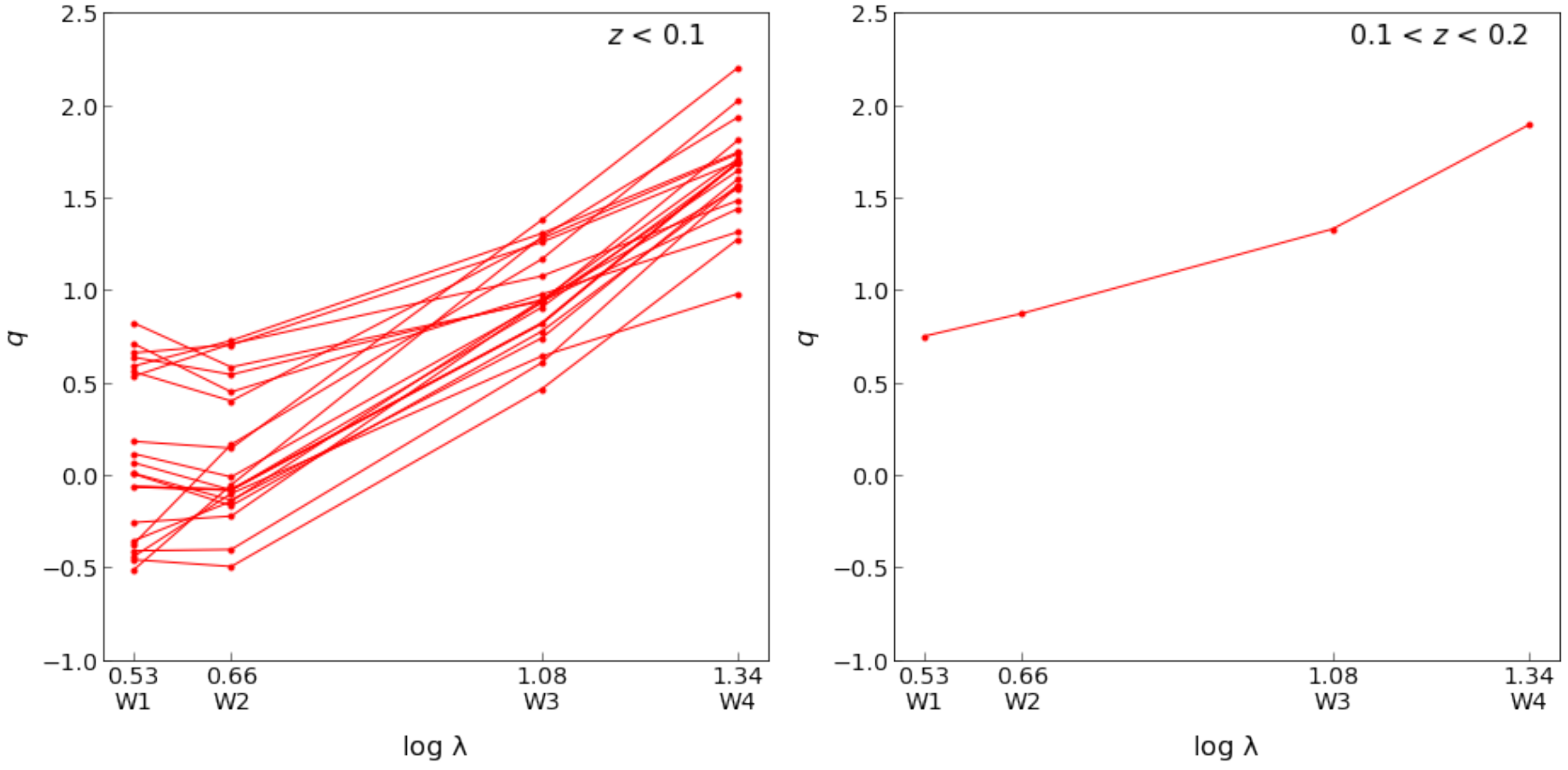

(a) (b)

**Figure 4.** The SED reflected by the various $q$-values for the (a) 20 ($z < 0.1$) and (b) 1 ($0.1 < z < 0.2$) high-$q_{150\ \mathrm{MHz}}^{22\ \mu\mathrm{m}}$ value sources.

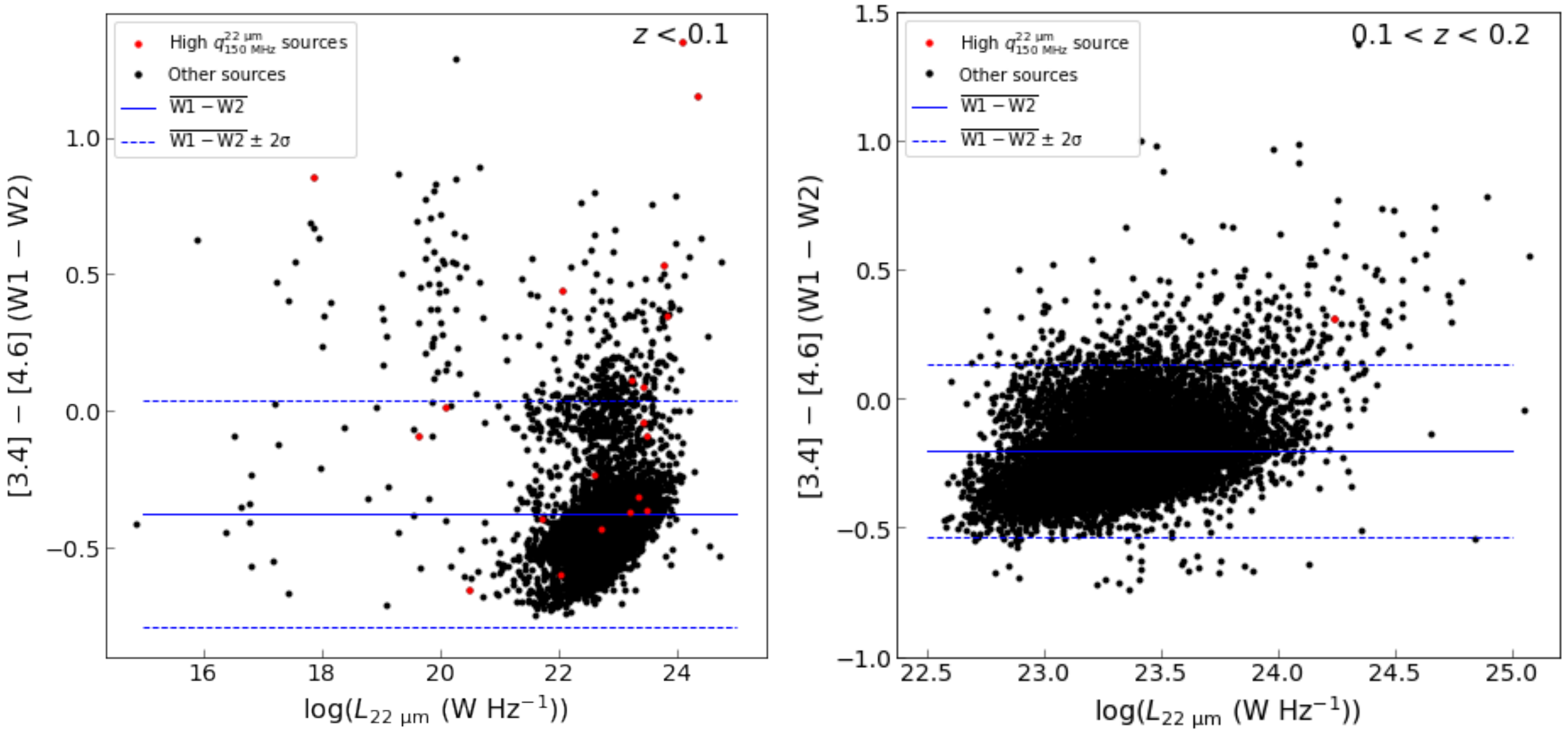

(a) (b)

**Figure 5.** The *WISE* MIR colour distribution of W1 - W2 versus 22 $\mu$m luminosity. The red points represent the (a) 20 ($z < 0.1$) and (b) 1 ($0.1 < z < 0.2$) high-$q_{150\ \mathrm{MHz}}^{22\ \mu\mathrm{m}}$ value sources. The blue lines in both panels denote the averaged values and $\pm\ 2\sigma$ away from it.

based on its location on the BPT diagram (Baldwin et al. 1981). Another source−−ILT J123803.78+461819.9 (SIMBAD identifier: SDSS J123803.75+461820.1) was classified as an emission-line galaxy in (Shim & Chary 2013; source ID: SDSS 588017109686943854) with H$\alpha$ line width measurement of $\gtrsim$ 500 Å. In addition, there are two sources (ILT J134649.72+542621.7 and ILT J145757.90+565323.8) that have little presence in the scientific literature. The faint optical emission of these two sources may be explained by the presence of dust that generates excessive MIR emission and causes extinction in the optical band but they deserve further study in the context of being Type II/III civilisations.

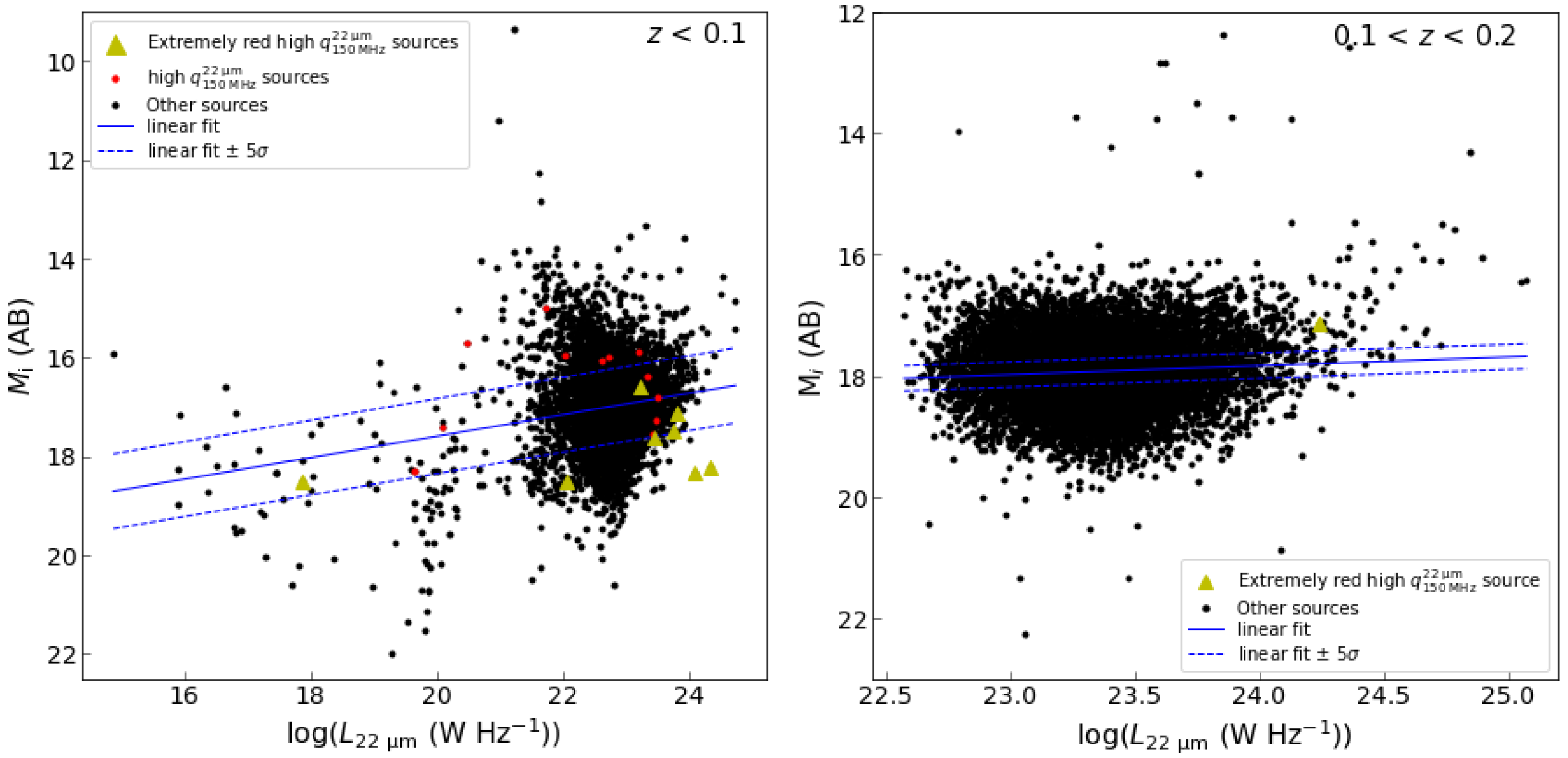

(a) (b)

**Figure 6.** The distributions of Pan-STARRS *i*-band forced aperture photometry magnitude. The red points represent the high-$q^{22\,\mu m}_{150\,\mathrm{MHz}}$ value sources. The yellow triangles represent extremely red sources within the high $q^{22\,\mu m}_{150\,\mathrm{MHz}}$ sample. The blue lines in both panels denote the least square linear fit over the whole sample and values of $\pm\,5\sigma$ away from it.

## 5 SUMMARY

We have created a sample of 16,367 sources selected from the LoTSS-DR1 value-added catalogue that includes radio and MIR flux density measurements, optical photomertry and photometric redshifts. We calculated $q^{22\,\mu m}_{150\,\mathrm{MHz}}$ values for 16,367 sources with $z < 0.2$, in an attempt to identify galaxies in the local Universe displaying the characteristics of energy intensive Kardasehv Type III civilisations. In particular, we looked for outliers in the IRC, searching for sources with $q^{22\,\mu m}_{150\,\mathrm{MHz}}$ > the linear fits +1 and $q^{22\,\mu m}_{150\,\mathrm{MHz}} > 0.25 + \log_{10}5$, one of the signatures expected of Type III civilisations generating significant amount of waste heat emission. 21 sources with such large $q$-values were found in the sample, corresponding to candidate Type III civilisations with their MIR radiation $\sim$ 10 times enhanced by waste heat are present in the sample.

We searched for identifiers for the 21 high-$q^{22\,\mu m}_{150\,\mathrm{MHz}}$ value sources from the SIMBAD catalogue −− 19 out of 21 of the sources are clearly identified as natural sources of cosmic emission, 2 sources have little presence in the scientific literature. The majority of the previously identified sources with high $q$-values are likely to be obscured AGN or star-forming ETG. All these high-$q^{22\,\mu m}_{150\,\mathrm{MHz}}$ value sources are radio-faint ($L_{150\,\mathrm{MHz}} \lesssim 10^{23}$ W Hz$^{-1}$) and have an SED similar to that of a typical spiral galaxy.

With the expectation of MIR excess caused by waste heat in this subsample of 21 sources, we looked for sources with SEDs peaking in the MIR. By doing a colour distribution analysis, within the 21 sources with high-$q^{22\,\mu m}_{150\,\mathrm{MHz}}$ value, we identified 9 sources with the most extreme MIR colours that is $2\sigma$ above the average. In addition to MIR-excess characteristics, we also looked for optical-faint features within the 9 extremely red high-$q$ sources. Amongst them, we found 4 outliers that showed significantly low optical/MIR luminosity ratio in the optical magnitude distribution, these sources have optical magnitudes that are $5\sigma$ fainter than the linear fitting of the whole sample. 2 of these 4 appear to be reasonably well studied, with characteristics that can be explained by natural astrophysical processes −− they are unlikely to be Type III civilisations. The other 2 sources (ILT J134649.72+542621.7 and ILT J145757.90+565323.8) warrant further observations. In addition, some of the sources in the two sub-samples with less extreme MIR colours are still poorly observed and understood.

The construction of a sample of high-$q^{22\,\mu m}_{150\,\mathrm{MHz}}$ values and very red, optically faint sources not only provides a good starting point for surveying advanced civilisations but also serves as an important input to studies of IRC modeling, AGN dust obscuration and galaxy evolution. We note that this very direct method of using extreme $q$-values to identify Kardashev TypeIII candidates can be extended to the full LoTSS survey area, and also to sources located at $z > 0.2$. This approach would generate a sample with an estimated 1 million sources (about 100 orders of magnitude greater than the sample presented here) and can place very strong constraints on the incidence of Type III civilisations in the Universe.

## ACKNOWLEDGEMENT

This work is supported by National Natural Science Foundation of China grant No. 11988101. This research made use of Astropy, a community-developed core Python package for Astronomy (Astropy Collaboration et al. 2013, 2018).

## DATA AVAILABILITY

The data underlying this article are available in the article.

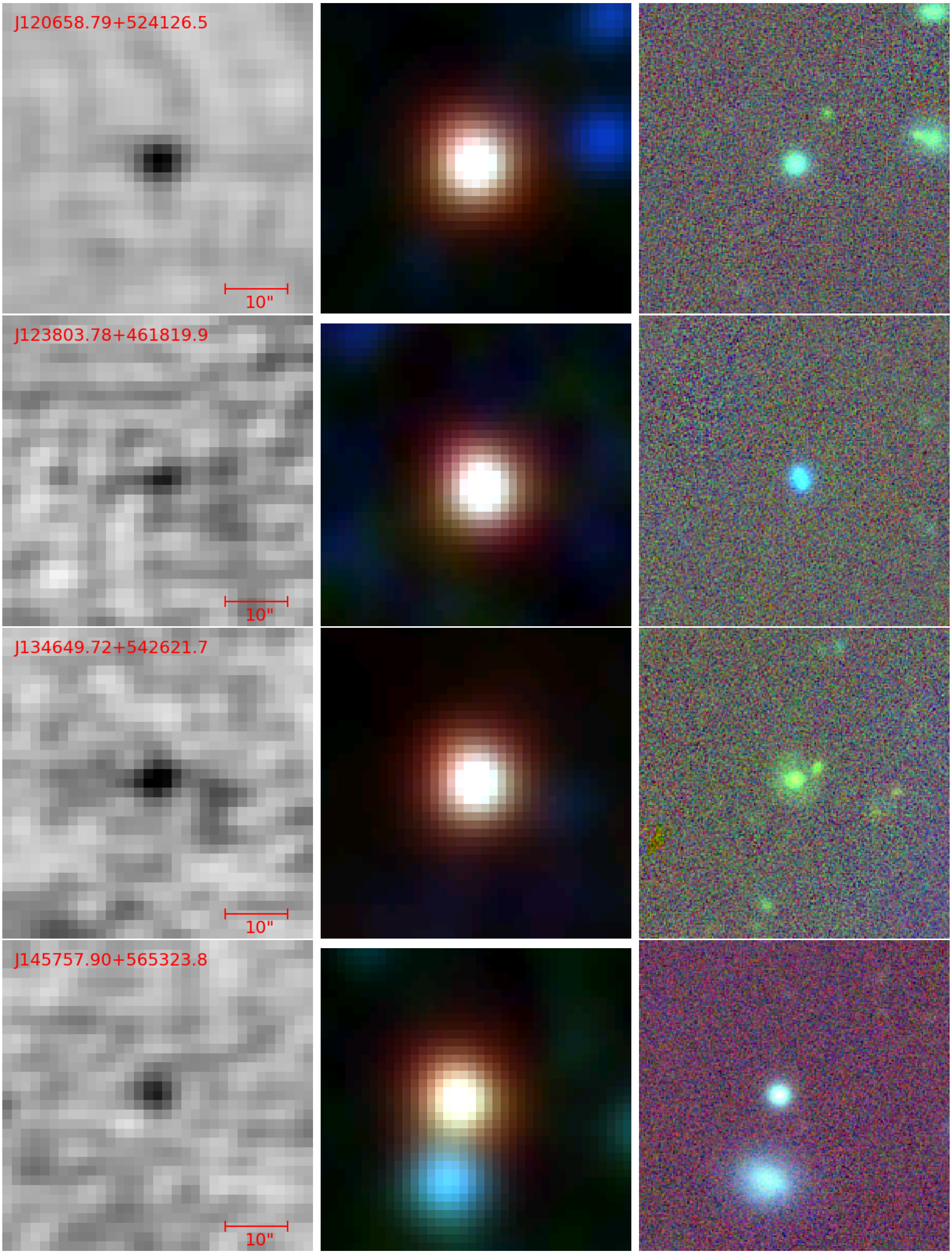

**Figure 7.** Multi-band images for the 4 very red, high $q_{150\,\mathrm{MHz}}^{22\,\mu\mathrm{m}}$ value, and optically faint sources. The LoTSS HBA 150 MHz radio images (left column) have a pixel size of 1.5″, the *WISE* (middle column) MIR multi-colour images are combined by 3.4 (blue), 4.6(green), and 12 (red) $\mu$m images, and have a pixel size of 1.5″, the Pan-STARRS optical optical images (right column) are combined by data from the Pan-STARRS y (red), i (green), g (blue) filters and have a pixel size of 0.25″. All the sub-figures have the same FoV of 50″ and are centered on the target coordinates drawn from the LoTSS DR1 catalogue.

Table 2: Information for the sources with high values of $q^{22\,\mu m}_{150\,MHz}$.

| LOFAR ID (J2000) [RA, DEC] (1) | $S_{150\,MHz}$ [mJy] (2) | $L_{150\,MHz}$ [W Hz$^{-1}$] (3) | $S_{22\,\mu m}$ [$\mu$ Jy] (4) | $L_{22\,\mu m}$ [W Hz$^{-1}$] (5) | $z$ (6) | mag$_{W1}$ (7) | mag$_{W2}$ (8) | mag$_{W3}$ (9) | mag$_{W4}$ (10) | mag$_i$ (11) | $q^{22\,\mu m}_{150\,MHz}$ (12) | Identifier (13) | Source Type (14) | Reference (15) |
|---|---|---|---|---|---|---|---|---|---|---|---|---|---|---|
| J104725.50+535338.5 | 0.516 | $5.399\times10^{21}$ | 25298.909 | $2.648\times10^{23}$ | 0.064 | 17.277 | 17.319 | 15.072 | 12.892 | 17.563 | 1.691 | LEDA 139444 | LIN AGN | 1 |
| J112637.84+513422.9 | 0.679 | $1.156\times10^{21}$ | 23979.250 | $4.080\times10^{22}$ | 0.026 | 15.228 | 15.463 | 14.453 | 12.950 | 16.060 | 1.548 | MCG+09-19-078 | Sy1/2 | 1; 2 |
| J112804.87+555630.8 | 1.645 | $1.412\times10^{21}$ | 60454.832 | $5.188\times10^{22}$ | 0.019 | 15.844 | 16.274 | 13.920 | 11.946 | 16.002 | 1.565 | 2MASX J11280486+5556303 | G | 1 |
| J115039.20+552112.7 | 0.640 | $1.509\times10^{19}$ | 13153.904 | $3.104\times10^{20}$ | 0.003 | 15.109 | 15.764 | 14.445 | 13.602 | 15.714 | 1.313 | NGC 3913 | G | 3 |
| J120522.51+502110.8 | 6.463 | $9.774\times10^{19}$ | 351854.592 | $5.321\times10^{21}$ | 0.003 | 12.979 | 13.374 | 11.181 | 10.034 | 15.006 | 1.736 | NGC 4085 | GiP | 4 |
| J120658.79+524126.5 | 1.318 | $2.532\times10^{22}$ | 112931.814 | $2.170\times10^{24}$ | 0.085 | 17.386 | 16.233 | 12.887 | 11.268 | 18.221 | 1.933 | SDSS J120658.79+524126.4 | Sy1 | 1 |
| J122615.87+482937.7 | 1.483 | $3.472\times10^{18}$ | 53937.728 | $1.263\times10^{20}$ | 0.001 | 16.996 | 16.979 | 14.456 | 12.070 | 17.419 | 1.561 | SDSS J122615.72+482938.6 | H2G | 5 |
| J123803.78+461819.9 | 0.451 | $1.187\times10^{22}$ | 47285.635 | $1.246\times10^{24}$ | 0.099 | 18.202 | 16.854 | 14.350 | 12.213 | 18.331 | 2.021 | SDSS J123803.75+461820.1 | EmG | 6 |
| J124505.63+561430.2 | 0.644 | $7.668\times10^{16}$ | 6134.263 | $7.300\times10^{17}$ | 0.0002 | 17.982 | 17.128 | 15.273 | 14.431 | 18.494 | 0.979 | J124505.6+561430 | QSO | 7 |
| J130708.14+461834.2 | 2.203 | $3.254\times10^{21}$ | 105655.486 | $1.560\times10^{23}$ | 0.025 | 15.376 | 15.746 | 13.277 | 11.340 | 15.884 | 1.681 | 2MASX J13070814+4618346 | AGN | 8 |
| J131048.89+544111.5 | 0.335 | $3.493\times10^{21}$ | 16538.928 | $1.722\times10^{23}$ | 0.064 | 15.937 | 15.825 | 14.441 | 13.354 | 16.602 | 1.693 | 2MASX J13104883+5441104 | Sy2 | 1 |
| J131459.85+511621.4 | 0.268 | $4.042\times10^{20}$ | 7364.117 | $1.110\times10^{22}$ | 0.025 | 15.774 | 16.372 | 15.482 | 14.232 | 15.964 | 1.439 | 2MASX J13145978+5116213 | LIN AGN | 1 |
| J131506.48+461409.7 | 0.453 | $1.180\times10^{22}$ | 25132.724 | $6.547\times10^{23}$ | 0.098 | 15.789 | 15.444 | 13.997 | 12.899 | 17.142 | 1.744 | 2MASS J13150651+4614101 | Sy1 | 9 |
| J131856.77+474609.6 | 1.259 | $5.009\times10^{21}$ | 55490.303 | $2.208\times10^{23}$ | 0.040 | 15.863 | 16.177 | 13.824 | 12.039 | 16.376 | 1.644 | 2MASX J13185676+4746092 | G | 1 |
| J134649.72+542621.7 | 0.469 | $3.741\times10^{20}$ | 14262.238 | $1.138\times10^{22}$ | 0.018 | 15.889 | 15.450 | 14.536 | 13.515 | 18.504 | 1.483 | | | |
| J140411.25+542519.6 | 1.166 | $2.358\times10^{18}$ | 21906.950 | $4.432\times10^{19}$ | 0.001 | 17.377 | 17.471 | 15.073 | 13.049 | 18.289 | 1.274 | SDSS J140411.24+542518.6 | H2G | 5 |
| J140956.92+545649.5 | 0.729 | $1.142\times10^{22}$ | 37215.466 | $5.832\times10^{23}$ | 0.077 | 17.634 | 17.099 | 14.380 | 12.473 | 17.467 | 1.708 | LEDA 2816082 | EmG/H2G | 5; 6 |
| J141424.94+465348.4 | 0.346 | $2.227\times10^{22}$ | 27003.192 | $1.740\times10^{24}$ | 0.150 | 15.679 | 15.371 | 14.230 | 12.821 | 17.147 | 1.893 | 2MASSJ14142489+4653485 | Sy1 | 10 |
| J143551.22+501118.2 | 0.498 | $7.584\times10^{21}$ | 19785.477 | $3.012\times10^{23}$ | 0.076 | 17.133 | 17.495 | 15.307 | 13.159 | 17.269 | 1.599 | SDSS J143551.27+501118.3 | G | 1 |
| J144632.38+490038.2 | 0.581 | $1.987\times10^{21}$ | 92241.223 | $3.153\times10^{23}$ | 0.037 | 16.536 | 16.627 | 13.546 | 11.488 | 16.820 | 2.201 | SDSS J144632.39+490038.0 | Comp | 11 |
| J145757.90+565323.8 | 0.806 | $4.262\times10^{21}$ | 51843.254 | $2.740\times10^{23}$ | 0.046 | 17.276 | 17.190 | 14.296 | 12.113 | 17.618 | 1.808 | | | |

(1) Source ID in the LoTSS value-added catalogue, (2) the total, integrated Stokes I 150 MHz radio flux density measured by the LOFAR HBA, (3) radio luminosity derived from LoTSS 150 MHz radio flux density, (4) *WISE* W4 profile-fit photometry flux, (5) calculated *WISE* 22$\mu$ luminosity, (6) Best available redshift estimate, (7)−(10) *WISE* W1−W4 profile-fit photometry magnitude, (11) Pan-STARSS forced aperture AB magnitude, (12) IRC parameter between the *WISE* 22 $\mu$m and LoTSS 150 MHz flux density, (13) Identifier archived in the SIMBAD catalogue, (14) Source type, G: Galaxy; Sy2: Seyfert 2 Galaxy; Sy1: Seyfert 1 Galaxy; Comp: Composite galaxy that likely have both star-forming and AGN activity; GiC: Galaxy in cluster of galaxies; LIN AGN: LINER-type AGN; QSO: Quasar object; EmG: Emission-line galaxy; X: X-ray source. (15) References; 1: Toba et al. (2014); 2: Dong et al. (2012); 3: Lavaux & Hudson (2011); 4: Crook et al. (2007); 5: Shirazi & Brinchmann (2012); 6: Shim & Chary (2013); 7: Allen et al. (2011); 8: Liu et al. (2011); 9: Oh et al. (2015); 10: Zhou et al. (2006); 11: Brinchmann et al. (2008);